\documentclass{ifacconf}

\usepackage[utf8]{inputenc}  
\usepackage[T1]{fontenc}     

\usepackage{graphicx}      
\usepackage{natbib}        
\usepackage{amsmath}	   
\usepackage{amssymb}	   
\usepackage{pgfplots}
\usepackage{booktabs}
\pgfplotsset{compat=1.14}
\usepackage{newfloat}
\DeclareFloatingEnvironment[placement={!t},name=Table]{mytable}
\DeclareFloatingEnvironment[placement={!t},name=Fig.]{myfigure}
\usepackage[format=plain, indention=0cm,width=\linewidth, labelfont=small]{caption}

\newcommand{\R}{\mathbb{R}}
\newcommand{\X}{\mathbb{R}^n}
\newcommand{\barX}{\bar{\mathbb{X}}_\infty}
\newcommand{\Xcl}{\mathbb{X}_\text{cl}^+}
\newcommand{\Z}{\mathbb{Z}}
\newcommand{\W}{\mathbb{W}}
\newcommand{\N}{\mathbb{N}}

\newcommand{\Q}{\mathbb{Q}}

\newcommand{\U}{\mathbb{R}^m}
\newcommand{\dint}{\,\mathrm{d}}
\newcommand{\zs}{z_\text{s}}
\newcommand{\vs}{v_\text{s}}
\DeclareMathOperator*{\argmin}{arg\,min}
\DeclareMathOperator{\interior}{int}
\newcommand{\norm}[1]{\left\| #1 \right\|}
\newcommand{\refeq}[2]{\overset{\makebox[0pt][c]{\scriptsize #1}}{#2}}
\newcommand{\Ni}[2]{\{#1..#2\}}

\theoremstyle{plain}
\newtheorem{theorem}{Theorem}
\newtheorem{defi}{Definition}
\newtheorem{rk}{Remark}
\newtheorem{ass}{Assumption}
\newtheorem{pro}{Proposition}
\newtheorem{coro}{Corollary}
\newtheorem{lemma}{Lemma}
\makeatletter
\let\origsection\section
\renewcommand\section{\@ifstar{\starsection}{\nostarsection}}

\newcommand\nostarsection[1]
{\sectionprelude\origsection{#1}\sectionpostlude}

\newcommand\starsection[1]
{\sectionprelude\origsection*{#1}\sectionpostlude}

\newcommand\sectionprelude{%
	\vspace{-.7em}
}

\newcommand\sectionpostlude{%
	\vspace{-1em}
}
\makeatother

\begin{document}
	\setlength{\abovedisplayshortskip}{0.7ex plus1ex minus1ex}
	\setlength{\abovedisplayskip}{0.7ex plus1ex minus1ex}
	\setlength{\belowdisplayshortskip}{0.7ex plus1ex minus1ex}
	\setlength{\belowdisplayskip}{0.7ex plus1ex minus1ex}
\begin{frontmatter}

\title{Robust Economic Model Predictive Control without Terminal Conditions\thanksref{footnoteinfo}} 

\thanks[footnoteinfo]{This work was supported by the German Research Foundation under Grants	GRK 2198/1, AL 316/12-2, and MU 3929/1-2.}

\author[IST]{Lukas Schwenkel}
\author[IST]{Johannes Köhler}
\author[IRT]{Matthias A. Müller}
\author[IST]{Frank Allgöwer}

\address[IST]{Institute for Systems Theory and Automatic Control, University of \makebox[1\linewidth]{\ \ Stuttgart,\,Germany,\,\{schwenkel,\,jokoehler,\,allgower\}@ist.uni-stuttgart.de}}
\address[IRT]{Institute of Automatic Control, Leibniz University, Hannover, Germany, mueller@irt.uni-hannover.de}

\begin{abstract}
	In this paper, a tube-based economic Model Predictive Control (MPC) scheme for systems subject to bounded disturbances is investigated that uses neither terminal costs nor terminal constraints.
	We provide robust guarantees on the closed-loop performance under suitable dissipativity and controllability conditions. 
	Furthermore, we prove practical convergence to an optimal robust control invariant set, as well as its practical stability under slightly stronger assumptions.
	Hence, this work extends the results from nominal economic MPC without terminal conditions to systems with bounded disturbances by using similar turnpike arguments and a properly modified stage cost.
	The results are discussed in a numerical example.
	
\end{abstract}


\end{frontmatter}

\section{Introduction}

In the past decade, economic Model Predictive Control (MPC) has emerged as an active field of research, see \citep{Faulwasser2018} for an overview. 
MPC schemes repeatedly solve finite horizon optimal control problems in a receding horizon fashion.
In economic MPC, the objective of this optimization problem is a general cost function, which often is given by some underlying economic considerations such as energy consumption, production amounts, etc., and does not need to be be positive definite with respect to a desired setpoint as it is standard in conventional MPC. 
A common technique to guarantee performance and stability of such MPC schemes is to add suitable terminal conditions to the optimal control problem.
These terminal conditions, however, can be difficult to design, lead to an unnecessary reduction of the region of attraction, and increase the computational burden such that, in practice, they are frequently dropped.
An explanation of why this practice is often successful was given by \cite{Gruene2013} and \cite{Gruene2014}; therein, the authors considered an economic MPC scheme without terminal conditions and provided practical stability as well as performance guarantees up to an error term vanishing with growing prediction horizons.

In the presence of uncertainties, \cite{Bayer2014} show that the performance of an economic MPC scheme can be significantly improved if the possible uncertainties are considered within the cost function of the optimization problem.
Optimal system operation, performance guarantees, and stability results of robust economic MPC have been considered thoroughly in recent years, see e.g., \citep{Bayer2016b}, \citep{Bayer2018}, and \citep{Dong2018}; all using terminal conditions.
In this work, we show that terminal conditions are often not needed to provide similar stability and performance results.
The results of \cite{Gruene2013} on the nominal case without terminal conditions do not apply to the robust setting, since the nominal closed-loop sequence in the proposed tube-based approach is no trajectory of the nominal system.
Hence, there are no theoretical closed-loop guarantees for such a scheme yet, only open-loop considerations are provided by \cite{Olshina2018}.

The main contribution of this paper is a thorough analysis of robust economic MPC without terminal conditions.
Assuming nominal dissipativity it is shown that the robust asymptotic average performance is no worse than the robust optimal steady-state (ROSS) performance up to an error vanishing with growing prediction horizons.
If further a robust dissipativity condition holds, we can show convergence and stability-like behavior of the nominal closed-loop sequence.
While convergence of the real closed-loop trajectory to a robust positive invariant set around the ROSS follows immediately, we need additional assumptions to show practical stability of this set.
The findings are illustrated and discussed in a numerical example.

\emph{Notations.} The set of continuous monotonically increasing functions $\alpha:\R_{\geq 0} \to \R_{\geq 0}$ with $\alpha(0)=0$ is denoted by $\mathcal K$, and by $\mathcal K_\infty$ if additionally $\alpha(x)\to \infty$ as $x\to \infty$. The set of continuous monotonically decreasing functions $\delta:\R_{\geq 0} \to \R_{\geq 0}$ with $\delta(t)\to 0$ as $t\to\infty$ is denoted $\mathcal L$. 
Further, $\beta \in \mathcal{KL}$, if $\beta(\cdot,t)\in\mathcal{K}$ and $\beta(x,\cdot) \in \mathcal{L}$ for all $x,t\geq 0$. 
For $x\in\R^n$, $\Omega\subset\R^n$, the point-to-set distance is denoted $\norm{x}_\Omega:= \inf_{y\in\Omega} \norm{x-y}$. The set of all integers in the interval $[a,b]$ is denoted by $\Ni{a}{b} := \Z \cap [a,b]$. The ball with radius $\rho$ and center $x\in\R^n$ is denoted by $B_\rho^n(x)$.
\section{Problem Setup}\label{sec:prob_setting}
The setup is mainly inherited from \cite{Bayer2018} except that the terminal cost and terminal constraints are omitted.
The system to be controlled is of the form
\begin{align}\label{eq:sys}
x(t+1)=f\big(x(t),u(t),w(t)\big), \quad x(0)=x_0
\end{align}
with $f:\X\times \U\times \W \to \X$ continuous, where $x(t)\in \X$ is the system state, $u(t) \in \U$ is the control input, and $w(t)\in \W \subseteq \R^q$ is an external disturbance. Moreover, the state and input constraints $(x(t),u(t))\in \Z\subseteq \X\times \U$ have to hold at all times $t\geq 0$.
\begin{ass}\label{ass:compact}
	The sets $\W$ and $\Z$ are compact, convex, have a nonempty interior, and $0\in\interior\W$.
\end{ass}
The control goal is to operate the system optimally with respect to a continuous (not necessarily positive definite) stage cost $L:\X\times \U \to \R$ such that the constraints are satisfied.
We follow a tube-based approach, which utilizes a continuous feedback parameterization 
\begin{align}\label{eq:pi}
	u(t) = \pi( x(t), v(t)),\quad \pi : \X \times \U \to \U
\end{align}
and invariant sets, compare e.g. \citep{Mayne2005} and \citep{Bayer2013}.
With this feedback we can rewrite the system dynamics, stage cost, and constraints as $f_\pi (x,v,w)=f(x, \pi( x, v),w)$, $L_\pi (x,v)= L ( x, \pi (x,v))$, and $\Z_\pi = \left\{ (x,v) \in \X\times \U \big| ( x, \pi (x,v) ) \in \Z\right\}$. 
We define the nominal system as
\begin{align}\label{eq:sys_nom}
z(t+1)&= f_\pi \big(z(t),v(t),0\big), \quad z(0)=z_0
\end{align}
and the error between the real and the nominal state as $e(t)=x(t)-z(t)$. 
This leads to the error dynamics
\begin{align}\label{eq:err_sys}
e(t+1)&= f_\pi \big(x(t),v(t),w(t)\big) -f_\pi \big(z(t),v(t),0\big).
\end{align}
\begin{defi}[\cite{Bayer2016b}] A compact set $\Omega\subseteq \R^n$ is \emph{robust control invariant} (RCI) for the error system \eqref{eq:err_sys} if there exists a feedback law \eqref{eq:pi} such that for all $x,z \in \X$ with $e=x-z \in \Omega$, $v\in\U$ with $(x,v) \in \Z_\pi$, and $w\in \W$ it holds that $e^+ = f_\pi (x,v,w) -f_\pi (z,v,0) \in \Omega$.
\end{defi}

A discussion how to find such an RCI set $\Omega$ is, e.g., provided in \cite{Bayer2016b}.
Henceforth, we assume its existence in the following.

\begin{ass}\label{ass:rci}
	There exists an RCI set $\Omega$ for the error system \eqref{eq:err_sys} such that $\Omega$ and 
	\begin{align}\label{eq:Zbar}
	\bar{\Z} &:= \left\{ (z,v) \in \Z_\pi \big| \forall \epsilon \in \Omega: (z+\epsilon,v)\in \Z_\pi
	\right\}
	\end{align} 
	are compact and have a nonempty interior.
\end{ass}
This assumption implies that the real system under the feedback \eqref{eq:pi} stays in a tube around the nominal state $x(t)\in\{z(t)\} \oplus \Omega$ for all times $t\geq 0$ if $e(0)\in\Omega$.
A further implication is $\bar \Z \oplus \Omega \subseteq \Z_\pi$, i.e., if the nominal system satisfies the tightened constraints $(z(t),v(t))\in\bar\Z$, then robust constraint satisfaction $(x(t),v(t))\in\Z_\pi$ follows.

\cite{Bayer2014} pointed out that for robust performance considerations it is not reasonable to take only the nominal stage cost into account and instead proposed to use an average cost over the possible real states
\begin{align}
L_\pi^{\text{int}}(z,v) := \int_{\Omega} L_\pi(z+\epsilon, v)\dint \epsilon.
\end{align}
As an alternative to the average, \cite{Bayer2016b} suggested a worst case stage cost
\begin{align} \label{eq:lmax}
L_\pi^\text{max}(z,v) := \max_{\epsilon \in \Omega} L_\pi(z+\epsilon,v).
\end{align}
Due to this selection, we generally denote the stage cost by $\ell$, which may be e.g., $L_\pi$, $L_\pi^\text{int}$, or $L_\pi^\text{max}$.
\begin{ass}\label{ass:lipschitz}
	The stage cost $\ell:\Z_\pi \to \R$ is \emph{Lipschitz continuous} on $\Z_\pi$ with Lipschitz constant $\kappa_\ell\in \R$, i.e. for all $(z_1,v_1),(z_2,v_2)\in\Z_\pi$ it holds
	\begin{align}\label{eq:lipschitz}
	|\ell (z_1,v_1) - \ell (z_2,v_2) | \leq \kappa_\ell \norm{(z_1,v_1)-(z_2,v_2)}.
	\end{align}
\end{ass}
The optimal control problem (OCP) that we propose to solve in each step of the tube-based MPC scheme is similar to \citep{Bayer2018}, however, without any terminal conditions, as these are often omitted in practice. 
Hence, the nominal initial state is also a decision variable - an idea that traces back to \citep{Mayne2005} and improves the performance.
Therefore, we will denote the OCP in two steps: First, given $z_0\in\X$, the nominal OCP is
\begin{subequations}\label{eq:ocp_nom}
\begin{flalign}
V_N \big(z_0\big)=\min_{z(\cdot),v(\cdot)}  &\,J_N \big(z_0, v(\cdot)\big)& \label{eq:ocp_opt_v}\\
\begin{split}
\text{s.t. }&z(k+1) = f_\pi\big(z(k),v(k), 0\big),\\ & z(0)=z_0, \end{split}\label{eq:ocp_sys}\hspace{-1cm}&\\
&\big( z(k),v(k) \big) \in \bar{\Z}, \label{eq:ocp_const}&\\
&\text{for all } k\in \Ni{0}{N-1}, \nonumber\\\label{eq:ocp_J}
\text{with} \qquad J_N \big(z_0, v(\cdot)\big) &= \sum_{k=0}^{N-1} \ell\big(z(k),v(k)\big),
\end{flalign}
\end{subequations}
with the minimizer $z_N^\star(\cdot;z_0)$, $v_N^\star(\cdot;z_0)$. Second, given $x(t)\in \X$, the robust OCP is
\begin{subequations}\label{eq:ocp}
\begin{align}\label{eq:ocp_opt_z0}
\mathcal V_N(x(t))= \min_{\makebox[0pt][c]{\scriptsize $z_0$}} &\,V_N(z_0)&\\\text{s.t.}\, &
\,x(t) \in \{z_0\}\oplus \Omega, \label{eq:ocp_init}&
\end{align}
\end{subequations}
where the minimizer is denoted by $z_0^\star(x(t))$. 
Further, we introduce the convenient notations $z_N^\star(\cdot|t):=z_N^\star(\cdot;z_0^\star(x(t)))$ and $v_N^\star(\cdot|t):=v_N^\star(\cdot;z_0^\star(x(t))$.
We can use the OCP \eqref{eq:ocp} in order to define the feedback law by the following MPC iteration. At each time instant $t$ we perform the steps
\begin{enumerate}
	\item Measure the current state $x(t)$ of the system,
	\item Solve OCP \eqref{eq:ocp} to get $v_N^\star (k|t)$, $k\in\Ni{0}{N-1}$,
	\item Apply the feedback $u(t)=\pi(x(t),v_N^\star(0|t))$,
\end{enumerate}
such that we obtain the closed-loop dynamics
\begin{align}\label{eq:closed_loop}
x(t+1)=f\big(x(t),\pi(x(t),v_N^\star(0|t)),w(t)\big).
\end{align}
The sets of all possible next nominal and real states of the MPC controlled closed-loop system are denoted by
\begin{flalign}
	\Z_\text{cl}^+(z)&:=\big\{z_0^\star (x_\text{cl}^+)\big| x_\text{cl}^+ \in \{f_\pi(z,v_N^\star (0;z),0)\}\oplus \Omega \big\},\\
	\Xcl(x)&:=\big\{f_\pi(x,v_N^\star (0;z_0^\star(x)),w)\big| w \in \W \big\}.
\end{flalign}
Moreover, we denote the set of all initial states for which there exists an infinite horizon admissible input by 
\begin{align}
\begin{split}
\barX &:= \{z\in\X| \exists\,(\tilde z, \tilde v):\N\to \bar \Z: \tilde z(0) = z \\
&\qquad \quad\wedge \tilde z (k+1) = f_\pi(\tilde z(k),\tilde v(k),0),  \forall k\geq 0 \}.
\end{split}
\end{align}

In our analysis, we will compare the performance to the one at the \emph{robust optimal steady state} (ROSS) defined by
\begin{align}
(\zs, \vs) := \argmin_{(z,v) \in \bar{\Z}, z=f_\pi(z,v,0)} \ell (z,v).
\end{align}
Note that $\zs$ (and $\vs$) depends on $\ell$. Thus, for the specific choices $\ell=L_\pi,$ $L_\pi^\mathrm{int}$, or $L_\pi^\mathrm{max}$ we write $\zs^L$, $\zs^\mathrm{int}$, or $\zs^\mathrm{max}$.
\begin{ass}\label{ass:ross_int}
	The ROSS lies in the interior of the constraint set, i.e. there is $\rho_3>0$ with $B_{\rho_3}^{m+n}(\zs,\vs) \subseteq \bar{\Z}$.
\end{ass}
For dissipative systems, \cite{Gruene2014} have shown that economic MPC schemes without terminal conditions are converging and satisfy  closed-loop performance bounds.
Compared to their work, we will assume a slightly stronger dissipation inequality that was used in \citep{Faulwasser2018} as well to simplify some of the later proofs, see Remark 3.1 therein for a discussion on different dissipation inequalities.

\begin{ass}\label{ass:strict_diss}
	The nominal system \eqref{eq:sys_nom} is \textit{strictly dissipative} on $\bar \Z$ with respect to the supply rate $s(z,v) = \ell(z,v) - \ell(\zs,\vs)$, i.e. there is $\alpha_\ell\in \mathcal K_\infty$ and a bounded storage function $\lambda: \X \to \R$ such that for all $(z,v)\in\bar{\Z}$ it holds
	\begin{align}\label{eq:strict_diss}
	\lambda \big(z^+\big) - \lambda(z) \leq s(z,v) - \alpha_\ell(\norm{(z,v)-(\zs,\vs)})
	\end{align}
	with $z^+=f_\pi(z,v,0)$.
\end{ass}

Further, we need some reachability and some local controllability of $\zs$, compare  \citep{Faulwasser2018} and \citep{Gruene2014}.
\begin{ass}\label{ass:exp_reach}
	The ROSS $(\zs,\vs)$ of system \eqref{eq:sys_nom} is \emph{exponentially reachable}, i.e. for all $z \in \barX$ there exists an infinite horizon admissible trajectory $(\zeta,\nu):\N\to\bar \Z$ and constants $c_1>0, \rho_1 \in [0,1)$ such that $\zeta(0)=z$ and for all $k\geq 0$ it holds $\zeta(k+1)=f_\pi(\zeta(k),\nu(k),0)$ and
	\begin{align}\label{eq:exp_reach}
	\norm{(\zeta(k),\nu(k))-(\zs,\vs)}\leq c_1\rho_1^k.
	\end{align}
\end{ass}
\begin{ass}\label{ass:local_ctrl}
	System \eqref{eq:sys_nom} is \emph{locally controllable} at $(\zs,\vs)$, i.e. there exist $\rho_2>0$, $M\in \N$, and $c_2>0$ such that for all $y,z \in B_{\rho_2}^n(z_s)$ there is $ \nu_y^z: \Ni{0}{M-1}\to\U$ with the corresponding nominal state trajectory $\zeta_y^z:\Ni{0}{M}\to\X$ satisfying \eqref{eq:sys_nom} and for all $k\in \Ni{0}{M-1}$
	\begin{subequations}
	\begin{align}
	 \zeta_y^z (0)=y,\quad &\quad \zeta_y^z (M)=z,\\
	\norm{\big(\zeta_y^z (k),\nu_{y}^z(k)\big) - (\zs,\vs)} &\leq c_2 \max_{\bar y \in \{y,z\}} \norm{\bar y-\zs}.
	\end{align}
	\end{subequations}
\end{ass}
\begin{rk}
	The constants $c_1,$ $c_2$, $\rho_1,$ $\rho_2,$ $\rho_3$, $\kappa_\ell$, $M$, and the $\mathcal K_\infty$ function $\alpha_\ell$ are unique throughout this paper. Whenever, for example, there appears $\rho_2$ in the manuscript it will refer to the local controllability Assumption \ref{ass:local_ctrl} without further referencing.
\end{rk}
\section{Performance Bound}\label{sec:perf_bound}
The goal of this section is to derive average performance bounds for the real closed-loop system controlled by the MPC scheme proposed in the previous section with prediction horizon $N$, i.e. to bound
\begin{align}
\mathcal J_T^\text{cl}(x(0),N) &= \frac 1 T \sum_{t=0}^{T-1} L_\pi \big( x(t), v_N^\star(0|t)\big), \\
\mathcal J_\infty^\text{cl}(x(0),N) &= \textstyle\limsup_{T\to \infty} \mathcal J_T^\text{cl} (x(0),N).
\end{align}
To this end, we will first derive bounds on the nominal closed-loop performance with the modified stage cost $\ell$
\begin{align}
 J_T^\text{cl}(z,N) &= \frac 1 T \sum_{t=0}^{T-1} \ell \big( z_N^\star(0|t), v_N^\star(0|t)\big), \\
J_\infty^\text{cl}(z,N) &= \textstyle\limsup_{T\to \infty} J_T^\text{cl} (z,N)
\end{align}
by exploiting the so-called \emph{turnpike property}, which follows from dissipativity and exponential reachability.
\begin{pro}\label{pro:turnpike}
	Let Assumptions \ref{ass:compact}--\ref{ass:lipschitz}, \ref{ass:strict_diss}, and \ref{ass:exp_reach} hold. Then the OCP \eqref{eq:ocp_nom} satisfies the \textit{turnpike property}, i.e. for all $N\in \N$, $\varepsilon>0$, and $z\in \barX$ it is $\# \Q_\varepsilon(N,z) \geq N-\frac{c_3}{\alpha_\ell(\varepsilon)}$
	with
	\begin{align}\label{eq:q_epsilon}
	\begin{split}
	\Q_\varepsilon(N,z) = \{k &\in \{0..N-1\}|\\& \norm{(z_N^\star(k;z),v_N^\star(k;z))-(\zs,\vs)} \leq \varepsilon \},
	\end{split}
	\end{align}
	where $c_3:=\kappa_\ell c_1(1-\rho_1)^{-1}+2\lambda_\text{max}$ and $\lambda_\text{max} = \sup_{(z,v)\in \bar \Z} |\lambda(z)|<\infty$.
\end{pro}
Whenever clear from context, we will write $\Q_\varepsilon=\Q_\varepsilon(N,z)$ in a slight abuse of notation.
A proof of Proposition \ref{pro:turnpike} can be found e.g. in \cite[Proposition 4.1]{Faulwasser2018}.
To derive a performance bound, we follow the steps in \citep{Gruene2013}.
However, the main difficulty with the tube-based setup is that the nominal closed-loop sequence is no longer a trajectory of the nominal system, since $z_N^\star(0|t)$ is a decision variable.
Still, we can modify Proposition 4.1 in \citep{Gruene2013} such that it extends to this setup. 
\begin{pro}\label{pro:perf_bound}
	Let Assumption \ref{ass:compact} and \ref{ass:rci} hold and assume there exists $N_0 \in \N$, $\delta_1,\delta_2 \in \mathcal L$ such that for all $N\geq N_0$ and for all $z \in \barX$ there exist $P \in \Ni 0 N$ and $\hat{v}_{N}(\cdot;z):\Ni{0}{N}\to\U$ with the corresponding nominal trajectory $\hat z_N(k;z)$ for $k\in\Ni 0 {N}$ satisfying the nominal dynamics \eqref{eq:sys_nom} and $(\hat z_N(k;z),\hat{v}_{N}(k;z))\in\bar \Z$ such that
	\begin{enumerate}
		\item[(i)] $\hat J_N (z) := \!\!\!\sum\limits_{k=0,\,k\neq P}^N\!\!\! \ell \big( \hat z_{N}
		(k;z), \hat v_{N}(k;z)\big) \leq V_N(z)+ \delta_1(N) $ \\
		\item[(ii)] $\ell \big( \hat z_N(P;z), \hat v_{N}(P;z)\big) \leq \ell (\zs,\vs) +\delta_2(N)$.
	\end{enumerate}
	Assume further that $z_N^\star (0|0) \in \barX$ implies $z_N^\star(1|t)\in \barX$ for all $t\geq 0$.
	Then for all $z=z_N^\star(0|0) \in \barX$ and all $T\in\N$ it holds with $\varepsilon_1=\delta_1+\delta_2\in\mathcal L$
	\begin{flalign}\label{eq:perf_transient}
	\begin{split}
	J_T^{\text{cl}} ( z,N ) &\leq \textstyle\frac{1}{T} V_{N}(z) - \frac{1}{T} V_{N} \big(z_N^\star(0|T)\big)+\ell (\zs,\vs)\\&\qquad\qquad\qquad\qquad\qquad\qquad\ \ + \varepsilon_1(N-1),\end{split}\hspace{-2em}&\\\label{eq:perf_asymptotic}
	J_\infty^{\text{cl}} ( z,N ) &\leq \ell (\zs,\vs) + \varepsilon_1(N-1).&
	\end{flalign}
\end{pro}
\begin{pf}
	The idea of the proof is to relate the closed-loop performance $J_T^\text{cl}(z,N)$ to the open-loop value functions $V_N(z)$ and bound them with (i), (ii). 
	The main difference to the proof of Proposition 4.1 in \citep{Gruene2013} is that due to the robust setup we have $z_N^\star (0|t) \neq z_N^\star (1|t-1)$.
	Nevertheless, we know from the optimality of $z_N^\star (0|t)$ that
	\begin{align}\label{eq:rel_nom_states}
	V_N \big(z_N^\star (0|t) \big)=\mathcal V_N(x(t)) \leq V_N \big(z_N^\star (1|t-1) \big)
	\end{align} 
	since $z_N^\star (1|t-1)$ satisfies \eqref{eq:ocp_init}, i.e. $x(t)\in \{z_N^\star (1|t-1)\} \oplus \Omega$.
	Step 1: We relate $J_T^\text{cl}(z,N)$ to $V_N(z)$ as follows
	\begin{align}\nonumber
	TJ_T^{\text{cl}} ( z,&N ) = \textstyle\sum_{\substack{t=0}}^{T-1} \ell \big( z_N^\star(0|t), v_N^\star(0|t)\big) \\\nonumber
	&= \textstyle\sum_{\substack{t=0}}^{T-1} \big(V_{N} \big( z_N^\star(0|t)\big)- V_{N-1} \big( z_N^\star(1|t)\big)\big)\\
	\begin{split}
	&\refeq{\eqref{eq:rel_nom_states}}{\leq} \textstyle\sum_{\substack{t=0}}^{T-2} \big(V_{N} \big( z_N^\star(1|t)\big)-V_{N-1} \big( z_N^\star(1|t)\big)\big)\\ &\qquad +  V_{N} ( z)- V_{N-1} \big( z_N^\star(1|T-1)\big).
	\end{split} \label{eq:JTcl_intermediate}
	\end{align}
	Step 2: Bound $V_N$ with (i) and (ii). For $y\in\barX$ it holds
	\begin{align}\nonumber
	V_{N}& ( y)-V_{N-1} ( y)\\\nonumber& \refeq{\text{(i)}}{\leq} J_N \big(y,\hat{ v}_{N-1}(\cdot;y)\big)-\hat J_{N-1} ( y)+\delta_1(N-1)\\\nonumber
	&=\ell \big(\hat z_{N-1}(P;y),\hat v_{N-1}(P;y)\big)+\delta_1(N-1)\\ \label{eq:VN_VN-1}
	&\refeq{\text{(ii)}}{\leq} \ell(\zs,\vs)+\delta_2(N-1)+\delta_1(N-1).
	\end{align}
	Step 3: Merge the two previous steps. If we insert \eqref{eq:VN_VN-1} with $y=z_N^\star (1|t)$, $t\in \Ni{0}{T-2}$ into \eqref{eq:JTcl_intermediate} and then use again \eqref{eq:VN_VN-1} with $y=z_N^\star (1|T-1)$ we obtain
	\begin{align*}
	TJ_T^{\text{cl}} ( z, N )\, &\refeq{(\ref{eq:JTcl_intermediate},\ref{eq:VN_VN-1})}{\leq} \, (T-1)\big(\ell(\zs,\vs)+\delta_2(N-1)+\delta_1(N-1)\big) \\ &\qquad + V_{N} ( z)- V_{N-1} \big( z_N^\star(1|T-1)\big)\\
	&\refeq{\eqref{eq:VN_VN-1}}{\leq} T \ell(\zs,\vs)+T\delta_2(N-1)+T\delta_1(N-1) \\ &\qquad + V_{N} ( z)- V_{N} \big( z_N^\star(1|T-1)\big)\\
	&\refeq{\eqref{eq:rel_nom_states}}{\leq}\, T\ell(\zs,\vs)+T\delta_2(N-1)+T\delta_1(N-1) \\ &\qquad + V_{N} ( z)- V_{N} \big( z_N^\star(0|T)\big).
	\end{align*}
	Since $\ell$ is continuous and $\bar{\Z}$ is compact, we can bound $N\ell_\text{max}\geq V_N(z) \geq N\ell_\text{min}$. Dividing by $T$ yields for $T\to\infty$
	\begin{flalign*}
	&&J_\infty^{\text{cl}} ( z, N ) \leq \ell(\zs,\vs)+\delta_2(N-1)+\delta_1(N-1).\qed
	\end{flalign*}
\end{pf}

In the following theorem, we will use the turnpike property and local controllability to show that indeed there exists a candidate solution satisfying the requirements of Proposition \ref{pro:perf_bound}; thereby proving recursive feasibility and the performance bounds \eqref{eq:perf_transient}, \eqref{eq:perf_asymptotic} for the proposed MPC scheme.

\begin{theorem}\label{theorem:perf_bound}
	Let Assumptions \ref{ass:compact}--\ref{ass:local_ctrl} hold. Then, there exists $N_0\in\N$ and $\varepsilon_1 \in \mathcal L$ such that for all $N\geq N_0$ the OCP \eqref{eq:ocp} is feasible for all times $t\geq 0$ if it is initially feasible with $z_0^\star(x(0))\in\barX$. Further, the performance estimates \eqref{eq:perf_transient} and \eqref{eq:perf_asymptotic} hold for all $z\in \barX$ and $T\geq 0$.
\end{theorem}
\begin{pf}
	To prove this result, we will use Proposition \ref{pro:perf_bound}.
	We start with showing recursive feasibility, which is assumed therein.
	To this end, we show $\Z_\text{cl}^+(z)\subseteq\barX$ for $z\in\barX$, which implies $z(t)\in\barX$ for all $t\geq 0$ if $z_0^\star(x(0))\in \barX$ and thus $x(t)\in\barX\oplus \Omega$ for all $t\geq 0$ such that the OCP \eqref{eq:ocp} is feasible for all $t\geq 0$.
	First, let $\delta \in \mathcal L$ be
	\begin{align*}
	\delta(N) = \alpha_\ell^{-1}\textstyle \left(\frac{c_3M}{N-M}\right)
	\end{align*}
	for $N>M$ and choose $N_0\in\N$ large enough such that $\delta(N_0)<\min\{\rho_2,c_2^{-1}\rho_3\}$.
	Then we know by the turnpike property from Proposition \ref{pro:turnpike} for $N\geq N_0$ and $z\in\barX$ that
	\begin{align}\label{eq:q_geq_N_M}
	\# \Q_{\delta(N)}(N,z) &\geq \textstyle N - \frac{c_3}{\alpha_\ell(\delta(N))} =N-\frac{N}{M}+1.
	\end{align}
	Without loss of generality we can assume $M\geq 2$ since $M=1$ implies that local controllability also holds with $M=2$. 
	Thus, there exists $P\in \Q_{\delta(N)}(N,z)$ with $P \geq 1$.
	Due to our choice of $N_0$ we know $\delta(N)<\min\{\rho_2,c_2^{-1}\rho_3\}$ for $N\geq N_0$ such that we can use the local controllability input to go from $z_N^\star (P;z)$ to $\zs$ since $\delta(N)<\rho_2$ and that this trajectory is feasible since $c_2\delta(N)<\rho_3$.
	This renders $z_N^\star(k;z)\in \barX$ for all $k \in \Ni {0}{P}$ and in particular for $k=1$.
	In the proof of the turnpike property of optimal solutions starting at $z\in\barX$ in Proposition 4.1 in \citep{Faulwasser2018} we see that $z\in\barX$ is only needed to give a constant upper bound on the value $V_N(z)$ via exponential reachability.
	The same upper bound holds for $V_N(z_N^\star(0;z_\text{cl}^+))$ since $z_N^\star(1;z)\in\barX$ and $V_N(z_N^\star(0;z_\text{cl}^+))\leq V_N(z_N^\star(1;z))$ due to optimality of $z_N^\star(0;z_\text{cl}^+)$ and the fact that $z_N^\star(1;z)$ is a feasible point for the OCP \eqref{eq:ocp} because the subsequent real system state $x_\text{cl}^+$ satisfies $x_\text{cl}^+ \in \{z_N^\star(1;z)\}\oplus\Omega $.
	Therefore, we conclude that optimal trajectories starting at $z_\text{cl}^+$ satisfy the turnpike property and as we have seen this implies that $\zs$ can be reached and thus $z_\text{cl}^+\in\barX$.
	The assumption $z_N^\star(1|t)\in\barX$ in Proposition \ref{pro:perf_bound} follows directly.
	
	Now let us construct a candidate solution of length $N+1$ that satisfies the assumptions (i) and (ii) of Proposition \ref{pro:perf_bound}.
	Combinatorial arguments and \eqref{eq:q_geq_N_M} yield that there are $M$ consecutive points $\Ni{P}{P+M-1}\subseteq \Q_{\delta(N)}(N,z)$, $P\in \N$ in a $\delta(N)$ neighborhood of $(\zs,\vs)$.
	To see this, assume for the sake of contradiction that no such $P$ exists, then there must be at least one element in each of the $\left\lfloor\frac{N}{M}\right\rfloor$ disjunct pieces of length $M$ in $\Ni 0 N$ that is not in $\Q_{\delta(N)}$, i.e. $\#\Q_{\delta(N)}\leq N- \left\lfloor\frac{N}{M}\right\rfloor$, which is a contradiction. 
	Due to $\Ni{P}{P+M-1}\subseteq \Q_{\delta(N)}(N,z)$, we can replace the middle piece $\Ni{P}{P+M-2}$ of $v_N^\star(\cdot;z)$ with the input $\nu_{z_1}^{z_2}(\cdot )$ from the local controllability for $z_1 = z_N^\star(P;z)$ and $z_2= z_N^\star(P+M-1;z)$, since both $z_1,z_2\in B_{\rho_2}^n(\zs)$ due to $P, P+M-1\in\Q_{\delta(N)}$ and $\delta(N)<\rho_2$, i.e.
	\begin{flalign}\label{eq:candidate_sol}
		\hat v_{N}(k;z) &= \begin{cases} v_N^\star(k;z) & \text{for } k\in \Ni 0 {P - 1} \\ \nu_{z_1}^{z_2}(k-P) & \text{for } k\in \Ni {P}{P+M-1} \\ v_N^\star(k-1;z) & \text{for } k\in \Ni{P+M}{N}. \end{cases}\hspace{-1cm}&
	\end{flalign}
	This input yields the open-loop nominal trajectory
	\begin{flalign}
	\hat z_N(k;z) &= \begin{cases} z_N^\star(k;z) & \text{for } k\in \Ni 0 {P - 1} \\
	 \zeta_{z_1}^{z_2}(k-P) & \text{for } k\in \Ni {P}{P+M-1} \\ 
	 z_N^\star(k-1;z) & \text{for } k\in \Ni {P+M}{N+1}, \end{cases}\hspace{-1cm}&
	\end{flalign}
	which has length $N+1$, and is feasible because of $B_{\rho_3}^{n+m}(\zs,\vs)\subseteq\bar{\Z}$ and $c_2\delta(N)<\rho_3$.
	To show the condition (i) of Proposition \ref{pro:perf_bound} we use
	\begin{align*}
	\hat J_N(&z)-V_N(z) = \!\!\textstyle\sum\limits_{\substack{k=0,\,k\neq P}}^N\!\! \ell \big( \hat z_N(k;z), \hat v_{N}(k;z)\big) - V_N(z)\\
	&=\textstyle\sum\limits_{k=1}^{M-1} \ell \big(\zeta_{z_1}^{z_2}(k), \nu_{z_1}^{z_2}(k)\big) -  \!\!\sum\limits_{k=P}^{P+M-2}\!\! \ell \big(z_N^\star(k;z),v_N^\star(k;z)\big)\\
	&\leq (M-1)(c_2+1)\kappa_\ell\delta(N)
	\end{align*}
	where $\kappa_\ell$ is the Lipschitz constant from Assumption \ref{ass:lipschitz}. Set $\delta_1(N) := (M-1)(c_2+1)\kappa_\ell \delta(N)$ to obtain (i). Further, we have due to the local controllability for $P$
	\begin{align*}
	\ell(\hat z_N(P;\!z),\hat v_{N}(P;\!z))=\ell(z_1,\nu_{z_1}^{z_2}(0)) &\leq \ell(\zs,\vs)\!+\! \kappa_\ell c_2 \delta(N).
	\end{align*}
	Set $\delta_2(N) := c_2\kappa_\ell \delta(N)$ to obtain (ii). Finally, apply Proposition \ref{pro:perf_bound} to conclude the proof. \qed
\end{pf}
The performance bounds in Theorem \ref{theorem:perf_bound} are only valid for stage cost $\ell$ of the nominal system.
The control goal, however, is to minimize the stage cost $L_\pi$ of the real system.
Depending on the choice of $\ell$ we can derive the following statements on the closed-loop average asymptotic performance of the real system, the transient performance bound \eqref{eq:perf_transient} can be shown analogue.
\begin{coro}\label{cor:bounds}
	Let Assumptions \ref{ass:compact}--\ref{ass:local_ctrl} hold. Then there exists $N_0\in\N$ and $\varepsilon_1 \in \mathcal L$ such that for all $N\geq N_0$ and all $x\in \barX\oplus \Omega$ it holds for $\ell = L^\mathrm{max}_\pi$
	\begin{flalign}\label{eq:perf_bound_x_Lmax}
	\mathcal J_\infty^{\text{cl}} ( x, N ) &\leq L^\mathrm{max}_\pi(\zs^\mathrm{max},\vs^\mathrm{max}) +\varepsilon_1(N-1)
	\end{flalign}
	and for $\ell = L_\pi$
	\begin{flalign}\label{eq:perf_bound_x_L}
	\mathcal J_\infty^{\text{cl}} ( x, N ) &\leq L_\pi(\zs^L,\vs^L)+ \kappa_{L_\pi} \max_{\epsilon \in \Omega} \norm{\epsilon} +\varepsilon_1(N-1) .\hspace{-1cm}&
	\end{flalign}
\end{coro}
\begin{pf}
	In the case of $\ell = L^\mathrm{max}_\pi$ we can directly estimate $L_\pi(x(t),v(t))\leq L^\text{max}_\pi(z(t),v(t))$ since $x(t)\in \{z(t)\}\oplus \Omega$ which yields $\mathcal J_\infty^{\text{cl}} ( x, N ) \leq J_\infty^{\text{cl}} ( z, N )$ for $x\in \{z\}\oplus \Omega$.
	In the case of $\ell = L_\pi$ we see that $L_\pi(x(t),v(t))\leq L_\pi(z(t),v(t))+ \kappa_{L_\pi} \max_{\epsilon \in \Omega} \norm{\epsilon}$ since $x(t)-z(t) \in \Omega$.
\end{pf}
\begin{rem}
	We can conclude that the robust performance guarantee \eqref{eq:perf_bound_x_Lmax} resulting from $\ell = L^\text{max}_\pi$ is better than \eqref{eq:perf_bound_x_L} resulting from $\ell = L_\pi$ due to $L^\text{max}_\pi(\zs^\text{max},\vs^\text{max}) \leq L_\pi^\text{max}(\zs^L,\vs^L) \leq L_\pi (\zs^L,\vs^L)+\kappa_{L_\pi}\max_{\epsilon\in\Omega}\norm{\epsilon}$.
	The improvement can be quite large as the motivation example in \citep{Bayer2014} shows.
	On the other hand, using $\ell = L^\text{max}_\pi$ is computationally more involved such that in some applications $L_\pi$ might be the only viable choice.
	For $\ell=L_\pi^\text{int}$ it is not easily possible to derive a bound on $\mathcal J_\infty^\mathrm{cl}(x,N)$, however, in this case \eqref{eq:perf_asymptotic} provides a bound on the performance averaged over the tube of possible closed-loop trajectories.
\end{rem}

\begin{rk}\label{rk:ass_diss_ind_constr}
In nominal economic MPC, the dissipativity Assumption \ref{ass:strict_diss} implies optimal operation at steady state, which means that no trajectory can achieve better performance than the steady state, see e.g. \citep{Faulwasser2018}.
Given the performance bound \eqref{eq:perf_asymptotic}, this yields to a convergence of $J_\infty^\text{cl}(z,N)$ to $\ell(\zs,\vs)$ up to $\varepsilon_1(N-1)$.
In the robust setting, however, this does not hold anymore since two consecutive closed-loop nominal states $z_N^\star (0|t)$ and $z_N^\star (0|t+1)$ do not satisfy the nominal system dynamics \eqref{eq:sys_nom}.
\cite{Bayer2018} have shown that robust optimal operation at steady state follows if the closed-loop nominal sequence obeys the dissipation inequality
\begin{align}\label{eq:diss_nominal_cl_seq}
\lambda\big(z_\text{cl}^+\big) \leq \lambda\big(z\big) + s\big(z,v_N^\star (0;z)\big)
\end{align}
for all $z\in \barX$ and for all $z_\text{cl}^+ \in \Z_\text{cl}^+(z)$, which is generically satisfied for $\Omega$-robustly dissipative\footnote{$\Omega$-robust dissipativity requires the dissipation inequality $\lambda(z^+)-\lambda(z)\leq s(z,v)$ to hold all feasible subsequent nominal states $z^+\in \Z_\text{cl}^+(z)$, see \cite{Bayer2018}.} setups.
This is a rather strong dissipativity formulation and often not satisfied, however, \eqref{eq:diss_nominal_cl_seq} can be enforced by the MPC design as \cite{Bayer2014} presented:
\begin{enumerate}
	\item[(i)] The inequality \eqref{eq:diss_nominal_cl_seq} can be implemented in the OCP \eqref{eq:ocp} to constrain the initial conditions $z_0$. Alternatively, \eqref{eq:diss_nominal_cl_seq} follows by the constraint
	\begin{align}\label{eq:diss_inducing_constr}
	\lambda(z_N^\star(0|t+1)) \leq\lambda(z_N^\star(1|t))
	\end{align}
	due to the nominal dissipativity of Assumption \ref{ass:strict_diss}.
	\item[(ii)] Fixing $z_0=z(t)=z_N^\star(1|t-1)$ in the OCP \eqref{eq:ocp} to be consistent with \eqref{eq:sys_nom} yields \eqref{eq:diss_nominal_cl_seq} due to Assumption \ref{ass:strict_diss}.
\end{enumerate}
Given \eqref{eq:diss_nominal_cl_seq}, it follows $J_T^\text{cl}(z,N)\geq \ell(\zs,\vs)$.
Together with the derived performance bound of Theorem \ref{theorem:perf_bound} this yields that the closed-loop average performance $J_T^\text{cl}(z,N)$ converges to the set $[\ell(\zs,\vs), \ell(\zs,\vs)+\varepsilon_1(N-1)]$ as $T\to\infty$ for $N\geq N_0$.
\end{rk}

\section{Practical Convergence}\label{sec:prac_stab_z}
In many applications, we do not want the closed loop to behave unpredictably and are interested in the optimal steady-state operation.
Therefore, we investigate in this section for what class of problems the nominal closed-loop sequence converges to $\zs$, which implicitly includes convergence of the real closed loop to the set $\{\zs\}\oplus \Omega$.
As the performance bound \eqref{eq:perf_asymptotic} comprises the error $\varepsilon_1(N-1)$, we only expect convergence to a neighborhood of $\zs$ that is shrinking with growing $N$, so-called practical convergence.
To this end, we assume \emph{strict} dissipativity of the nominal closed-loop sequence compared to dissipativity \eqref{eq:diss_nominal_cl_seq}, which is required for convergence of the performance.
\begin{ass}\label{ass:strict_diss_ind_constraint}
	Let Assumption \ref{ass:strict_diss} hold. For all $z\in \barX$ and $z_\text{cl}^+ \in \Z_\text{cl}^+(z)$ the following dissipation inequality holds
	\begin{align}\label{eq:strict_diss_inducing_constr}
	\begin{split}
	\lambda\big(z_\text{cl}^+\big) &\leq \lambda(z) + s\big(z,v_N^\star (0;z)\big)\\&\qquad -\alpha_\ell(\norm{\big(z,v_N^\star (0;z)\big)-(\zs,\vs)}).
	\end{split}
	\end{align}
\end{ass}
If this assumption is not satisfied, it can be induced by constraints in the OCP as discussed in Remark \ref{rk:ass_diss_ind_constr}.
 
\cite{Gruene2014} have shown practical asymptotic stability for nominal economic MPC without terminal conditions by establishing the rotated value function 
\begin{align}\label{eq:rot_ocp}
\tilde V_N(z_0) &= \min_{v(\cdot),z(\cdot),\,\text{s.t.\,(\ref{eq:ocp_sys},\ref{eq:ocp_const})}} \sum_{k=0}^{N-1} \tilde{\ell}(z(k),v(k))
\end{align}
with the minimizer $\tilde{z}_N^\star(\cdot;z_0)$, $\tilde{v}_N^\star(\cdot;z_0)$ as practical Lyapunov function, where the rotated stage cost is defined by
\begin{align}\label{eq:rot_cost}
\tilde{\ell}(z,v)&= \ell(z,v)-\ell(\zs,\vs) +\lambda(z) - \lambda(f_\pi(z,v,0)).
\end{align}
This rotated value function is also crucial in the stability proof of economic MPC with terminal conditions, see \citep{Amrit2011} for the nominal and \citep{Bayer2014} for the robost case.
However, an essential difference to these works when omitting terminal conditions is that the minimizer of the rotated OCP is generally not identical to the minimizer of the original OCP.

\begin{rk}\label{rk:rot_turnpike}
If the strict dissipativity condition (Assumption \ref{ass:strict_diss}) holds for the original OCP \eqref{eq:ocp_nom}, then it also holds for the rotated OCP \eqref{eq:rot_ocp} with respect to the rotated supply $\tilde s (z,v)=\tilde \ell(z,v) - \tilde \ell(\zs,\vs) = \tilde \ell(z,v)$. 
This follows immediately from \eqref{eq:strict_diss} with the rotated storage $\tilde{\lambda} = 0$ and $\tilde \alpha_\ell=\alpha_\ell$.
Thus, we can show the turnpike property as in Proposition \ref{pro:turnpike} for the rotated OCP with the same constant $c_3$ and the same function $\alpha_\ell$.
\end{rk}

We will heavily exploit the similarity of optimal solutions following from the turnpike property.
In particular, we will exploit that two solutions of the same OCP have end pieces with similar costs and that two solutions with the same initial condition but one of the original and one of the rotated OCP have start pieces with similar costs as stated in the following Lemma, which is inspired by Lemma 7.3 and Lemma 7.5 in \citep{Gruene2013}.

\begin{lemma}\label{lem:turnpike_trajectory_similarities}
Consider the two OCPs
\begin{flalign}\label{eq:ocp_general_i}
V_{N}^i(z_0)&= \min_{v(\cdot),z(\cdot),\,\text{s.t.\,(\ref{eq:ocp_sys},\ref{eq:ocp_const})}} J_N(z_0,v(\cdot)) + F^i(z(N))\hspace{-1cm}&
\end{flalign}
differing in their terminal costs $F^i:\X\to\R$, for $i\in\{1,2\}$ and let the trajectory of the corresponding minimizer be denoted $(z_N^{i\star}(\cdot;z_0), v_{N}^{i\star}(\cdot;z_0))$. 
Let Assumption \ref{ass:compact}--\ref{ass:ross_int} and \ref{ass:local_ctrl} be satisfied. Then, for $c_3=M(c_2+1)\kappa_\ell$, $\rho_4=\min\{\rho_2, c_2^{-1} \rho_3\}$, and $i,j\in\{1,2\}$ the following statements hold:
\begin{enumerate}
	\item[(i)] For all $y,z\in\barX$, all $\delta\in(0,\rho_4]$, and all $P\in \N$ with $\Ni P {P+M}\subseteq\Q^i_{\delta}(N,z)\cap \Q^i_{\delta}(N,y)$ it holds
\end{enumerate}
	\begin{align}\label{eq:tp_sim_same_ocp}
	\begin{split}
	\big|V^i_{N-P}(z_{N}^{i\star}(P;y)) - V^i_{N-P}(z_{N}^{i\star}(P;z))\big| &\leq c_3 \delta .
	\end{split}
	\end{align}
\begin{enumerate}
	\item[(ii)] For all $z\in\barX$, all $\delta\in(0,\rho_4]$, and all $P\in\N$ with $\Ni P{P+M}\subseteq \Q^i_{\delta}(N,z)\cap \Q^j_{\delta}(N,z)$ it holds 
	\begin{align}\label{eq:tp_sim_same_init}
	\big|J_P(z,v_{N}^{i\star}(\cdot;z)) - J_P(z,v_{N}^{j\star}(\cdot;z))\big| \leq c_3 \delta .
	\end{align}
\end{enumerate}
The set $\Q_\delta^i(N,z)$ is defined analogue to \eqref{eq:q_epsilon} but for the optimal sequence of OCP \eqref{eq:ocp_general_i}. 
\end{lemma}
\begin{pf}
	To show (i) we use the local controllability Assumption \ref{ass:local_ctrl} to go from $z_{N}^{i\star}(P;y)\in B_\delta^n (\zs)$ to $z_{N}^{i\star}(P+M;z)\in B_\delta^n (\zs)$ and the Lipschitz continuity of $\ell$ to obtain
	\begin{align*}
		V^i_{N-P}(z_{N}^{i\star}(P;y)) &\leq V^i_{N-P-M}(z_{N}^{i\star}(P+M;z)) \\&\quad+ M \ell(\zs,\vs) + Mc_2\kappa_\ell \delta\\
		& \leq V^i_{N-P}(z_{N}^{i\star}(P;z)) + M(c_2+1)\kappa_\ell \delta
	\end{align*}
	where the last inequality holds due to $\Ni{P}{P+M}\subseteq \Q^i_{\delta}(N,z)$.
	Interchanging $y$ and $z$ yields the absolute value.
	For verifying (ii) we use the local controllability to go from $z_{N}^{j\star}(P;z)\in B_\delta (\zs)^n$ to $z_{N}^{i\star}(P+M;z)\in B_\delta (\zs)^n$ and the Lipschitz continuity of $\ell$ to obtain
	\begin{align*}
	&V^i_{N}(z)\leq J_P(z,v_{N}^{j\star}(\cdot;z)) + M\ell(\zs,\vs) + Mc_2\kappa_\ell \delta\\&\qquad +V^i_{N-P-M}(z_{N}^{i\star}(P+M;z)) \\
	&\leq J_P(z,v_{N}^{j\star}(\cdot;z))+V^i_{N-P}(z_{N}^{i\star}(P;z))  + M(c_2+1)\kappa_\ell \delta 
	\end{align*}
	where the last inequality holds since $\Ni P{P+M}\subseteq \Q^i_{\delta}(N,z)$.
	Noting that
	\begin{align*}
	J_P(z,v_{N}^{i\star}(\cdot;z)) = V^i_{N}(z)  - V^i_{N-P}\big(z_{N}^{i\star}(P;z)\big),
	\end{align*}
	leads to the desired inequality without the absolute value, but interchanging $i$ and $j$ completes the proof.\qed
\end{pf}
\begin{rk}\label{rk:ocps}
We can recover the original OCP \eqref{eq:ocp_nom} $V^1_{N}(z)=V_N(z)$ with $F^1(z)=0$. 
The lemma can also be used for the rotated OCP \eqref{eq:rot_ocp} with $F^2(z) = -\lambda(z)$ if one considers that $V^2_{N}(z)=\tilde V_N(z)-\lambda(z)+N\ell(\zs,\vs)$. The minimizers $\tilde {v}_N^\star(\cdot;z)$ and $v_N^{2\star}(\cdot;z)$ are the same since the cost functions only differ by a constant depending on $z$.
\end{rk}
We can use the rotated value function $\tilde V_N$ to show convergence of the nominal closed-loop sequence up to an error term vanishing with growing $N$ using Lyapunov arguments. 
\begin{theorem}\label{theorem:pract_stab_nom}
	Let Assumptions \ref{ass:compact}--\ref{ass:strict_diss_ind_constraint} hold. Then there exist $N_1\in\N$ and $\varepsilon_2\in\mathcal L$ such that for all $N\geq N_1$ there is $\beta \in \mathcal{KL}$ with
	\begin{flalign}\label{eq:practical_convergence}
	\norm{z_N^\star(0|t)-\zs} \leq \max \left\{\beta\left(\norm{z_N^\star(0|0)-\zs},t\right),\varepsilon_2(N) \right\}\hspace{-1cm}&&
	\end{flalign}
	 for any sequence $z_N^\star(0|t+1)\in \Z_\text{cl}^+(z_N^\star(0|t))$, $t\geq 0$ with initial value $z_N^\star(0|0)\in \barX$.
\end{theorem}
\begin{pf}
	We will show that $\tilde V_N$ is a practical Lyapunov function w.r.t. $\delta_3(N)$, which means that there exist $\alpha_1,\alpha_2,\alpha_3\in\mathcal{K}_\infty$ such that for all $z\in \barX$ it holds
	\begin{align}\label{eq:lyap_candidate}
		\alpha_1 (\norm{z-\zs}) \leq \tilde V_N (z) &\leq \alpha_2(\norm{z-\zs})
	\end{align}
	and for all $z_\text{cl}^+ \in \Z_\text{cl}^+(z)$ it holds
	\begin{align}\label{eq:lyap_decrease}
		\tilde V_N (z_\text{cl}^+) &\leq \tilde V_N (z) - \alpha_3(\norm{z-\zs}) +\delta_3(N).
	\end{align}
	Then, we can conclude for all sequences $z_N^\star (0|t+1) \in \Z_\text{cl}^+(z_N^\star(0|t))\subseteq \bar{\X}_\infty$ convergence in the sense of \eqref{eq:practical_convergence} w.r.t. $\varepsilon_2(N)=\alpha_1^{-1}(\alpha_2(\alpha_3^{-1}(\delta_3(N)))+\delta_3(N))$ as stated e.g. in \citep{Gruene2014}.
	Due to Assumption \ref{ass:strict_diss} it holds
	\begin{align*}
	\tilde V_N(z)&=\min_{\substack{v(\cdot),z(\cdot),\,\text{s.t.\,(\ref{eq:ocp_sys},\ref{eq:ocp_const})}}} \textstyle\sum_{k=0}^{N-1} \tilde \ell (z(k),v(k))\\
	&\geq \min_{\substack{v(\cdot),z(\cdot),\\\text{s.t.\,(\ref{eq:ocp_sys},\ref{eq:ocp_const})}}} \textstyle\sum_{k=0}^{N-1} \alpha_\ell(\norm{(z(k),v(k))-(\zs,\vs)})\\
	&\geq \alpha_\ell(\norm{z-\zs})=:\alpha_1(\norm{z-\zs}).
	\end{align*}
	To construct an upper bound we distinguish $z\in B_{\rho_4}^n (\zs)$ and $z\not\in B_{\rho_4}^n (\zs)$, $\rho_4=\min\{\rho_2, c_2^{-1} \rho_3\}$. 
	For $z\in B_{\rho_4}^n (\zs)$ we can use the local controllability since $\rho_4\leq\rho_2$ to construct a candidate solution that is feasible since $\rho_4\leq c_2^{-1} \rho_3$ and steers the system from $z$ to $\zs$ in $M$ steps.
	Since $(\zs,\vs)$ is a steady state, the candidate solution can stay there with $\tilde \ell (\zs,\vs)=0$ for $N-2M$ steps and finally the local controllability can be used again to steer from $\zs$ to $z$ in the last $M$ steps to cancel out the storage function in the rotated costs
	\begin{align*}
	\tilde V(z) &\leq \textstyle\sum_{k=0}^{M-1} \tilde\ell \big(\zeta_{z}^{\zs} (k),\nu_{z}^{\zs}(k)\big) +\sum_{k=0}^{M-1} \tilde\ell \big(\zeta_{\zs}^{z} (k),\nu_{\zs}^{z}(k)\big) \\
	&\leq 2M\kappa_\ell c_2 \norm{z-\zs}.
	\end{align*} 
	For $z\not\in B_{\rho_4} (\zs)$ we define $M'\in\N$ such that $c_1\rho_1^{M'}<\rho_4$, which is the number of time steps needed for the exponential reachability to reach $B_{\rho_4}(\zs)$. 
	In this neighborhood of $\zs$, we can reach $\zs$ within $M$ steps without violating the constraints.
	Thus, with $K=M+M'$ we have
	\begin{align*}
	\tilde V_N (z) &\leq K \ell_\text{max} + 2 \lambda_\text{max} - K  \ell(\zs,\vs) =:C \leq \frac C {\rho_4} \norm{z-\zs}.
	\end{align*}
	In summary, we can conclude for both cases
	\begin{align*}
	\tilde V_N (z) &\leq \max\left\{2M\kappa_\ell c_2, \frac C {\rho_4} \right\} \norm{z-\zs} =: \alpha_2(\norm{z-\zs}).
	\end{align*}
	As next step, we verify the decrease \eqref{eq:lyap_decrease} up to $\delta(N)$, $\delta\in\mathcal L$ of the practical Lyapunov candidate $\tilde V_N$. To this end, we want to show that there exists $\delta_3 \in \mathcal L$ with
	\begin{align}\label{eq:lyap_decrease_intermediate}
	\begin{split}
	\tilde V_N (z)- &\tilde V_N (z_\text{cl}^+) + \delta_3(N) \geq s(z,v_N^\star(0;z))-\lambda(z_\text{cl}^+)+\lambda(z),
	\end{split}
	\end{align}
	for all $z_\text{cl}^+ \in \Z_\text{cl}^+(z)$ such that with \eqref{eq:strict_diss_inducing_constr} we directly obtain \eqref{eq:lyap_decrease} with $\alpha_3=\alpha_\ell$ and $\delta=\delta_3(N)$. 
	We will exploit the turnpike property of the five trajectories $z_N^\star(\cdot;z)$, $\tilde{z}_N^\star(\cdot;z)$, $z_N^\star(\cdot;z_\text{cl}^+)$, $\tilde{z}_N^\star(\cdot;z_\text{cl}^+)$, and $ z_N^\star(\cdot;z_\text{ol}^+)$ with $z_\text{ol}^+=f_\pi(z,v_N^\star (0;z),0)$.
	Hence, in order to use Lemma \ref{lem:turnpike_trajectory_similarities}, we have to choose $N_1\in\N$ large enough, such that for all $N \geq N_1$ there exists $P\in \N$ with $\Ni P{P+M} \subseteq \Q'$, where
	\begin{align}\label{eq:turnpike_times_5}
	\Q':=\Q_\delta(N, z_\text{ol}^+) \cap\! \textstyle\bigcap_{y\in\{z,z_\text{cl}^+\}} \!
	\big(\Q_\delta(N, y)\cap \tilde\Q_\delta(N, y)\big).
	\end{align}
	Thus, if we choose $\delta=\delta(N)$, i.e. $\delta\in\mathcal{L}$, as
	\begin{align*}
	\delta(N) = \textstyle \alpha^{-1}_\ell \left(\frac{5c_3(M+1)}{N-M-1}\right)
	\end{align*}
	then there exists $P$ such that $\Ni P{P+M} \subseteq \Q'$ holds. 
	To see that, assume for the sake of contradiction that no such $P$ exists, then there must be at least one element in each of the $\left\lfloor\frac{N}{M+1}\right\rfloor$ pieces of length $M+1$ in $\Ni 0 N$ that is not in $\Q'$, i.e. $\#\Q'\leq N- \left\lfloor\frac{N}{M+1}\right\rfloor$.
	Further, we know due to the turnpike property for each of the five trajectories that
	\begin{align*}
	\# \Q_{\delta(N)} (N,y) \geq \textstyle N-\frac{N}{5(M+1)}+\frac 1 5
	\end{align*}
	and thus $\# \Q'\geq N- \frac{N}{M+1} +1$,
	which is a contradiction.
	Now choose $N_1\in\N$ large enough such that $\delta(N_1)<\rho_4$ to ensure feasibility.
	Having established the requirements of Lemma  \ref{lem:turnpike_trajectory_similarities}, we can use it in the sense of Remark \ref{rk:ocps} to begin showing \eqref{eq:lyap_decrease_intermediate}
	\begin{align*}
	\Delta\ &\!:= \tilde V_N(z_\text{cl}^+)-\tilde V_N(z) +\lambda(z)-\lambda(z_\text{cl}^+)\\&\refeq{\eqref{eq:tp_sim_same_ocp}}{\leq}J_P\big(z_\text{cl}^+,\tilde {v }_N^\star(\cdot;z_\text{cl}^+)\big)-J_P\big(z,\tilde {v }_N^\star(\cdot;z)\big) +c_3 \delta(N) \\
	&\refeq{2x\eqref{eq:tp_sim_same_init}}{\leq}J_P\big(z_\text{cl}^+, {v }_N^\star(\cdot;z_\text{cl}^+)\big)-J_P\big(z, {v }_N^\star(\cdot;z)\big) +3c_3 \delta(N).
	\end{align*}
	Since $z_\text{cl}^+=\argmin_{\{y|x_\text{cl}^+-y\in \Omega\}} V_N(y)$ and $x_\text{cl}^+-z_\text{ol}^+\in \Omega$ we know that $V_N(z_\text{cl}^+) \leq V_N(z_\text{ol}^+)$. Using \eqref{eq:tp_sim_same_ocp}, this implies
	\begin{align*}
	J_P\big(z_\text{cl}^+, {v }_N^\star(\cdot;z_\text{cl}^+)\big) &\leq J_P\big(z_\text{cl}^+, {v }_N^\star(\cdot;z_\text{ol}^+)\big) + c_3 \delta(N).
	\end{align*}
	Thus, we arrive at
	\begin{align*}
	\Delta&\leq J_P\big(z_\text{ol}^+, {v }_N^\star(\cdot;z_\text{ol}^+)\big)-J_P\big(z, {v }_N^\star(\cdot;z)\big) +4c_3 \delta(N).
	\end{align*}
	In order to relate $J_P\big(z_\text{ol}^+, {v }_N^\star(\cdot;z_\text{ol}^+)\big)$ with $J_{P+1}\big(z, {v }_N^\star(\cdot;z)\big)$ we have to use the candidate solution $\hat v_{N}(\cdot;z)$ from \eqref{eq:candidate_sol} for $P+1$ instead of $P$ to obtain in a first step
	\begin{align*}
	V_N(z_\text{ol}^+) &\leq J_N\big(z_\text{ol}^+,\big(\hat v_{N}(k;z)\big)_{k\in \Ni 1 N}\big)\\
	&\leq J_{P}\big(z_\text{ol}^+,\big(v_{N}^\star (k;z)\big)_{k\in \Ni 1{P+1}}\big)+ M \ell(\zs,\vs)\\&\qquad +c_3\delta(N)+V_{N-P-M}(z_N^\star(P+M;z)) \\
	&\leq J_{P+1}(z,v_{N}^\star (\cdot;z))-\ell(z,v_N^\star(0;z))+V_{N}(z)\\&\qquad -J_P(z,v_N^\star(\cdot;z))+2c_3\delta(N)\\
	&\leq \ell(\zs,\vs)-\ell(z,v_N^\star(0;z))+V_{N}(z)+3c_3\delta(N).
	\end{align*}
	Now we can use \eqref{eq:tp_sim_same_ocp} to get rid of the end pieces of the trajectories and obtain
	\begin{align*}
	J_P(z_\text{ol}^+,v_N^\star (\cdot;z_\text{ol}^+)) &\leq \ell(\zs,\vs)-\ell(z,v_N^\star(0;z))\\&\quad+J_{P}(z,v_N^\star(\cdot;z))+4c_3\delta(N),
	\end{align*}
	which finally leads to \eqref{eq:lyap_decrease_intermediate} 
	\begin{align*}
	\Delta&\leq \ell(\zs,\vs)-\ell(z,v_N^\star(0;z)) +8c_3 \delta(N)
	\end{align*}
	with $\delta_3(N) = 8c_3\delta(N)$ and completes the proof. \qed 
\end{pf}
\begin{rk}
	Due to the convergence of the nominal closed-loop sequence \eqref{eq:practical_convergence}, the real trajectory satisfies 
	\begin{align}\nonumber
	\norm{x(t)}_{\{\zs\}\oplus \Omega} &\leq 
	\underbrace{\norm{x(t)}_{\{z(t)\}\oplus \Omega}}_{=0,\text{ due to }\Omega \text{ RCI}} + \norm {z(t)-\zs} \\
	\leq \max&\left\{\beta(\norm{z_N^\star(0|0)-\zs},t), \varepsilon_2(N)\right\}, \label{eq:traj_conv}
	\end{align}
	which proves that the trajectory $x(t)$ converges to $\{\zs\}\oplus \Omega$ up to $\varepsilon_2(N)$ as $t\to\infty$.
\end{rk}
\section{Practical asymptotic stability}\label{sec:prac_stab_x}
The convergence of $x(t)$ to $\{\zs\}\oplus \Omega$ up to $\varepsilon_2(N)$ does \emph{not} yet prove practical asymptotic stability of $\{\zs\}\oplus \Omega$, since $\norm{z_N^\star(0|0)-\zs}$ is not necessarily proportional to $\norm{x(0)}_{\{\zs\}\oplus \Omega}$.
Nevertheless, we can prove practical asymptotic stability with respect to $\varepsilon_3(N)\geq \varepsilon_2(N)$ if we assume that the storage function is continuous and has a maximum at $\zs$.
\begin{ass}\label{ass:lambda_continuous_zs_max}
	There exists $\alpha_\lambda\in\mathcal K_\infty$ such that for all $z \in \X$ the storage function $\lambda$ satisfies the inequalities
	\begin{align}\label{eq:zs_lambda_max}
	\lambda(\zs)&\geq \lambda(z),\\\label{eq:zs_lambda_cont}
	\lambda(\zs)-\lambda(z) &\leq \alpha_\lambda (\norm{z-\zs}).
	\end{align}
\end{ass}
This Assumption is similar to the one required in \citep{Bayer2017} to show asymptotic stability of robust economic MPC with terminal conditions and is needed to ensure that there are no directions from which $\zs$ can be reached cheaper than it costs to stay at $z(0)=\zs$. 
If this were the case, even $x(0)=\zs$ could cause $z(0)\neq\zs$ and thus $x(t)$ could leave $\{\zs\}\oplus \Omega$.
However, with Assumption \ref{ass:lambda_continuous_zs_max} we can conclude closed-loop practical stability of the real system.
\begin{theorem}\label{thm:stability}
	Let Assumptions \ref{ass:compact}--\ref{ass:lambda_continuous_zs_max} hold. Then there exist $N_1\in\N$ and a function $\varepsilon_3\in\mathcal L$ such that for all $N\geq N_1$ the set $\{\zs\}\oplus \Omega$ is practically asymptotically stable w.r.t. $\varepsilon_3(N)$ under the closed-loop dynamics \eqref{eq:closed_loop}, i.e. for all $x(0) \in \barX \oplus \Omega$ it holds
	\begin{align}\label{eq:practical_stability}
	\norm{x(t)}_{\{\zs\}\oplus\Omega} \leq \max \big\{\beta\big(\norm{x(0)}_{\{\zs\}\oplus\Omega},t\big),\varepsilon_3(N) \big\}
	\end{align}
	for any trajectory $x(t+1)\in \Xcl(x(t))$, $t\geq 0$.
\end{theorem}
\begin{pf}
	For this proof, we use a more general version of the Lyapunov characterization of practical asymptotic stability that also allows for an error in the upper bound of the Lyapunov function and modify it for stability of a set. 
	This is possible, since the proof for estimating the sequence $\norm{x(t)-\zs}$ does not change compared to a sequence $\norm{x(t)}_{\{\zs\}\oplus \Omega}$.
	In particular, we define the rotated value function for the real state as
	\begin{align}\label{eq:rot_cost_x}
	\tilde {\mathcal V}_N(x) = \tilde V_N(z_0^\star(x))
	\end{align}
	where $z_0^\star(x)$ is the minimizer of \eqref{eq:ocp}. 
	We show for all $x\in\barX \oplus \Omega$ and all $x_\text{cl}^+\in\Xcl(x)$ that
	\begin{flalign} \label{eq:lyap_candidate_x}
	&\alpha_1 (\norm{x}_{\{\zs\}\oplus \Omega}) \leq \tilde {\mathcal V}_N (x) \leq \alpha_4(\norm{x}_{\{\zs\}\oplus \Omega})+\delta_4(N)\hspace{-1cm}&\\
	&\tilde {\mathcal V}_N (x_\text{cl}^+) \leq \tilde {\mathcal V}_N (x) - \alpha_3(\norm{x}_{\{\zs\}\oplus \Omega}) +\delta_3(N) \label{eq:lyap_decrease_x}
	\end{flalign}
	holds to obtain practical asymptotic stability of the set $\{\zs\}\oplus\Omega$ for the real closed-loop trajectory w.r.t. $\varepsilon_3=\alpha_1^{-1} \circ (\alpha_4 \circ (\alpha_3^{-1} \circ \delta_3 +\delta_3)+\delta_4 + \delta_3)$, see \cite[Proposition 4.3]{Faulwasser2018}.
	For the first inequality of \eqref{eq:lyap_candidate_x} it is easy to see that
	\begin{align*}
	\tilde {\mathcal V}_N(x) = \tilde V_N(z_0^\star(x)) \geq \alpha_1(\norm{z_0^\star(x)-\zs}) \geq \alpha_1 (\norm{x}_{\{\zs\}\oplus \Omega}).
	\end{align*}
	To show the second inequality of \eqref{eq:lyap_candidate_x}, we have to distinguish $x\in B_{\rho_4}^n(\zs)\oplus \Omega $ from $x \notin B_{\rho_4}(\zs)\oplus \Omega$. 
	In the first case we define $z_\text{min}= \argmin_{\{z|x-z \in\Omega\}} \norm{z-\zs}$, for which a feasible trajectory exists since $\norm{z_\mathrm{min} -\zs}<\rho_4\leq\rho_2$ allows to construct a trajectory with the help of the local controllability that steers in $M$ steps to $\zs$ and stays there, where the feasibility follows from $\rho_4\leq c_2^{-1}\rho_3$.
	We will use Lemma \ref{lem:turnpike_trajectory_similarities} to bound $\tilde {\mathcal V}_N(x) = \tilde V_N(z_0^\star (x))$ w.r.t. $\tilde V_N (z_\mathrm{min})$.
	Note that the trajectories $z_N^\star(\cdot;z_0^\star(x))$, $\tilde{z}_N^\star(\cdot;z_0^\star(x))$, $z_N^\star(\cdot;z_\mathrm{min})$, and $\tilde{ z}_N^\star(\cdot;z_\mathrm{min})$ satisfy the turnpike property and that $N_1$ was chosen large enough such that any five trajectories share at least $M+1$ consecutive points in an $\delta(N)=\delta_3(N)/(8c_3)$ neighborhood of $\zs$.
	In particular, this holds for these four trajectories, thus, there exists $P\in\N$ such that $\Ni P{P+M} \subseteq \Q'$.
	Therefore, it is
	\begin{align} \nonumber J_P(z_0^\star(x),\tilde{v}_N^\star(\cdot;z_0^\star(x))) &\refeq{\eqref{eq:tp_sim_same_init}}{\leq} J_P(z_0^\star(x),{v}_N^\star(\cdot;z_0^\star(x)))+ c_3\delta(N)\\\nonumber
	&\refeq{(\ref{eq:tp_sim_same_ocp},\ref{eq:z_min_opt})\quad}{\leq} J_P(z_\mathrm{min},{ v}_N^\star(\cdot;z_\mathrm{min})) + 2c_3\delta(N)\\\label{eq:Jp_z_zmin}
	&\refeq{\eqref{eq:tp_sim_same_init}}{\leq} J_P(z_\mathrm{min},\tilde{ v}_N^\star(\cdot;z_\mathrm{min})) + 3c_3\delta(N)
	\end{align}
	where we used the optimality of $z_0^\star (x)$
	\begin{align}\label{eq:z_min_opt}
	V_N(z_0^\star (x))\leq V_N(z_\mathrm{min})
	\end{align}
	at the marked inequality. Thus, we have with $V_N^2$ from Remark \ref{rk:ocps} and due to Assumption \ref{ass:lambda_continuous_zs_max}
	\begin{align*} \tilde V_N (z_0^\star&(x)) = V_N^2 (z_0^\star(x)) + \lambda(z_0^\star (x)) - N\ell (\zs,\vs)\\
	&\refeq{(\ref{eq:tp_sim_same_ocp},\ref{eq:Jp_z_zmin})\quad }{\leq} J_P(z_\mathrm{min}, \tilde{ v}_N^\star(\cdot;z_\mathrm{min}))+\lambda(z_0^\star (x))- N\ell (\zs,\vs) \\&\qquad + V_{N-P}^2(\tilde z_N^\star(P;z_\mathrm{min})) + 4c_3\delta(N)\\
	&= V_{N}^2(z_\text{min})+\lambda(z_0^\star (x))- N\ell (\zs,\vs)+ 4c_3\delta(N) \\
	&\refeq{\eqref{eq:zs_lambda_max}}{\leq} \tilde V_N(z_\mathrm{min})+\lambda(\zs) - \lambda(z_\mathrm{min}) + 4 c_3\delta(N)\\
	&\refeq{\eqref{eq:zs_lambda_cont}}{\leq} \alpha_2(\norm{z_\mathrm{min}-\zs})+\alpha_\lambda(\norm{z_\mathrm{min} - \zs}) + 4c_3 \delta(N).
	\end{align*}
	Defining $\alpha_4=\alpha_2+\alpha_\lambda$ and $\delta_4=4c_3\delta(N)$ it follows
	\begin{align*}
	\tilde {\mathcal V}_N(x) = \tilde V_N(z_0^\star (x))&\leq \alpha_4(\norm{x}_{\{\zs\}\oplus \Omega})+\delta_4(N).
	\end{align*}
	For $x \notin B_{\rho_4}^n(\zs)\oplus \Omega$ we can show 
	\begin{align*}
	\tilde {\mathcal V}_N(x) = \tilde V_N(\tilde z_0^\star (x)) \leq \frac {C}{\rho_4} \norm{x}_{\{\zs\}\oplus\Omega}\leq \alpha_2(\norm{x}_{\{\zs\}\oplus\Omega})
	\end{align*}
	analogue to this case in the proof of Theorem \ref{theorem:pract_stab_nom}.
	Condition \eqref{eq:lyap_candidate_x} follows due to $\alpha_2\leq \alpha_4$.
	To show \eqref{eq:lyap_decrease_x}, we note that
	\begin{align*}
	\tilde {\mathcal V}_N (x_\text{cl}^+) &= \tilde V_N(z_0^\star(x_\text{cl}^+)) = \tilde V_N (z_\text{cl}^+) 
	\\& \refeq{\eqref{eq:lyap_decrease}}{\leq} \tilde V_N (z_0^\star(x))- \alpha_3(\norm{x}_{\{\zs\}\oplus \Omega})+\delta_3(N)
	\\& = \tilde{\mathcal{V}}_N (x)- \alpha_3(\norm{x}_{\{\zs\}\oplus \Omega})+ \delta_3(N).
	\end{align*}
	This establishes that $\tilde{\mathcal V}_N$ is a practical Lyapunov function and thus $\{\zs\}\oplus\Omega$ is practically asymptotically stable with respect to $\varepsilon_3(N)$. \qed
\end{pf}
\begin{rk}
	The fact that $\varepsilon_3(N)\geq \varepsilon_2(N)$ results from asymptotic stability (guaranteed up to $\varepsilon_3$) being a stronger property than asymptotic convergence (guaranteed up to $\varepsilon_2$), since it also includes bounds for the transient behavior.
	This transient behavior, however, depends crucially on the choice of the first nominal state $z_0^\star(x(0))$, $x(0)-z_0^\star(x(0))\in \Omega$, such that we needed the additional Assumption \ref{ass:lambda_continuous_zs_max} to ensure that $z_0^\star(x(0))$ is close to $\zs$ whenever $x(0)$ is close to $\{\zs\}\oplus\Omega$.
\end{rk}
\begin{rk}
	The statement of Theorem \ref{thm:stability} does also hold under Assumptions \ref{ass:compact}--\ref{ass:strict_diss_ind_constraint}, if one modifies the MPC such that at time $0$ the nominal initial state is not chosen as $z_N^\star(0|0)=z_0^\star(x(0))$ the minimizer of \eqref{eq:ocp_nom} but as $z_N^\star(0|0)=z_\text{min}(x(0)):=\argmin_{\{z |x(0)-z\in \Omega\}} \norm{z-\zs}$. This choice guarantees that $\norm{z_N^\star(0|0)-\zs} = \norm{x(0)}_{\{\zs\}\oplus \Omega}$ holds and thus \eqref{eq:traj_conv} implies \eqref{eq:practical_stability} even with $\varepsilon_2(N)$ instead of $\varepsilon_3(N)$. 
\end{rk}
\section{Numerical example}
Consider the scalar economic growth model from \cite{Gruene2014} with dynamics
\begin{align}
x(k+1)=u(k) + w(k)
\end{align}
and the stage cost $L(x,u)=-\ln (Ax^\alpha-u)$, with $A=5$ and $\alpha=0.34$, where we introduced an additive disturbance $w(k)\in\W=[-1,1]$.
The state and input constraints are $\Z=[0,10]\times [0.1,5]$. 
Since the error dynamic is stable, we do not need the pre-stabilizing feedback $\pi$, i.e. $\pi(x,v)=v$ such that $f=f_\pi$, $\Z=\Z_\pi$ and $L=L_\pi$.
Since the system is static, the RCI set $\Omega$ can be chosen as $\Omega=\W$.
Lipschitz continuity of the cost function $L$ on the feasible set $\Z$ is not given in this example as $L$ is not even defined everywhere in $\Z$.
To fix this, we relax the logarithm $\ln(\xi)$ for $\xi<0.001$ to a parabola, that smoothly continues $\ln$ at $0.001$.
As discussed in \citep{Gruene2014}, the problem is (nominally) strictly dissipative with the linear storage function $\lambda(x)=0.2306 x$.
Thus, we conclude that Assumptions \ref{ass:compact}--\ref{ass:local_ctrl} hold such that Corollary \ref{cor:bounds} provides the bound \eqref{eq:perf_bound_x_Lmax} on the asymptotic average performance.

Indeed, we can verify \eqref{eq:perf_bound_x_Lmax} in the simulation results shown in Fig. \ref{fig:exmp} with prediction horizon $N=10$, initial condition $x(0)=\zs^\text{max}$, and $w(k)$ sampled uniformly from $\W$.
The real closed-loop trajectory (yellow dots) is always in cheaper regions than $L_\pi^\mathrm{max} (\zs^\mathrm{max},\vs^\mathrm{max}) = -1.2$.
However, since the nominal state sequence (yellow circles) does not converge to $\zs^\text{max}$, we conclude that Assumption \ref{ass:strict_diss_ind_constraint} is not satisfied.
Thus, we implement the dissipativity inducing constraint \eqref{eq:diss_inducing_constr} in the OCP formulation as discussed in Remark \ref{rk:ass_diss_ind_constr}, (ii).
We can see in Fig. \ref{fig:exmp} that this forces the sequence of nominal states to converge to (stay at) $\zs^\mathrm{max}$.
Surprisingly, the performance values in Table \ref{tab:perf_measures} indicate that this additional constraint in the optimization improves the average cost in closed loop although the MPC controller has less freedom to choose the control action.
Further, we can see  in Fig. \ref{fig:exmp} and Table \ref{tab:perf_measures} that the running cost is on average over time and in the best case better for $\ell = L_\pi$ and $\ell = L_\pi^\mathrm{int}$.
Finally, we verified that using $\ell=L_\pi^\mathrm{max}$ leads to the best worst case performance and using $\ell=L_\pi^\mathrm{int}$ leads to the best average performance, although in this example, the difference to $\ell=L_\pi$ is almost negligible.
For shorter horizons, we observed practical convergence up to an perceptible error $\varepsilon(N)$ as in \citep{Gruene2014}.  
By increasing the horizon to $N=10$ this error gets negligible, such that in Fig. \ref{fig:exmp} the limit point of the nominal closed-loop sequence and $(\zs,\vs)$ are not distinguishable.

\begin{myfigure}
	\input{sections/level_plot.tex}\vspace{-3mm}
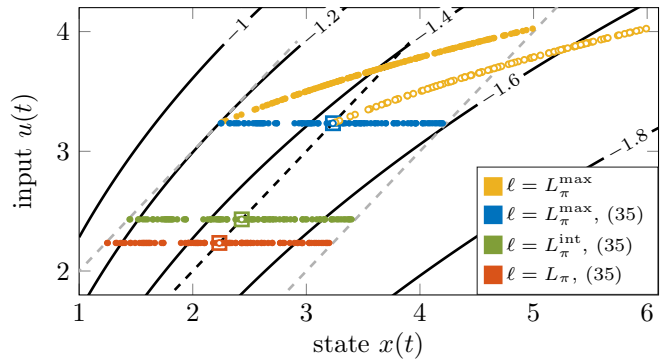
	\caption{\small The state-input domain with levels of the stage cost $L$ (black, solid), the manifold of steady states $x=f(x,u)$ (black, dashed), and the $\Omega$-tube around it (gray, dashed). As we can see, the optimal steady state (square $\boldsymbol\square$) depends on the choice of $\ell$. The real closed-loop trajectory (dots $\bullet$) and the nominal closed-loop sequence (circles $\boldsymbol\circ$) converge to $\{\zs\}\oplus \Omega \times \{\vs\}$ and $(\zs,\vs)$, respectively, only if we explicitly enforce robust dissipativity \eqref{eq:diss_inducing_constr} in the OCP, as the simulations $t\in[0,100]$ starting at the optimal steady state $\zs$ show.}\label{fig:exmp}
\end{myfigure}
\begin{mytable}
	\centering
	\small
	\begin{tabular}{ccc}
		\toprule
		Approach & Worst case & Average \\[-0.1cm] \midrule
		$\ell=L_\pi^\text{max}$ & $-1.209$ & $-1.403$ \\
		$\ell=L_\pi^\text{max}$, \eqref{eq:diss_inducing_constr} & $-1.205$ & $-1.427$ \\
		$\ell=L_\pi^\text{int}$, \eqref{eq:diss_inducing_constr} & $-1.168$ & $-1.447$ \\
		$\ell=L_\pi$, \eqref{eq:diss_inducing_constr} & $-1.143$ & $-1.445$ \\[-0.1cm]
		\bottomrule\\[-0.6cm]
	\end{tabular}\caption[width=90cm]{\small
			Mean and maximum of the real stage cost $L(x(t),u(t))$ along the closed-loop trajectory over $t\in[0,2000]$ for different choices of the stage cost $\ell$ used in the OCP with and without enforcing \eqref{eq:diss_inducing_constr}.
		}\label{tab:perf_measures}
\end{mytable}

\section{Conclusion}
In this work, we have verified that robust economic MPC can be implemented without terminal conditions if the system satisfies suitable (nominal) controllability and dissipativity conditions.
We have derived performance bounds not only on the nominal closed-loop sequence w.r.t. the modified stage cost, but also for the perturbed closed-loop trajectory w.r.t. the real stage cost.
If the system satisfies a robust dissipativity condition, or if dissipativity of the nominal closed-loop sequence is enforced with an additional constraint in the OCP, then the real closed loop practically converges to the set $\{\zs\}\oplus\Omega$, where the system is robust optimally operated. The error $\varepsilon_2(N)$ vanishes with increasing prediction horizons $N$.
To summarize, we have extended the economic MPC framework from \cite{Gruene2013} such that it can handle disturbances and we have shown that the robust EMPC scheme by \cite{Bayer2014} can be implemented without terminal cost and constraints under suitable assumptions.

\let\oldthebibliography=\thebibliography
\let\endoldthebibliography=\endthebibliography
\renewenvironment{thebibliography}[1]{%
	\begin{oldthebibliography}{#1}%
		\small
		\setlength{\parskip}{0ex}%
		\setlength{\itemsep}{0ex}%
	}%
	{%
	\end{oldthebibliography}%
}
\bibliography{my_bib}

\begin{thebibliography}{12}
\providecommand{\natexlab}[1]{#1}
\providecommand{\url}[1]{\texttt{#1}}
\providecommand{\urlprefix}{URL }
\expandafter\ifx\csname urlstyle\endcsname\relax
  \providecommand{\doi}[1]{doi:\discretionary{}{}{}#1}\else
  \providecommand{\doi}{doi:\discretionary{}{}{}\begingroup
  \urlstyle{rm}\Url}\fi

\bibitem[{Amrit et~al.(2011)Amrit, Rawlings, and Angeli}]{Amrit2011}
Amrit, R., Rawlings, J.B., and Angeli, D. (2011).
\newblock Economic optimization using model predictive control with a terminal
  cost.
\newblock \emph{Annual Reviews in Control}, 35(2), 178 -- 186.

\bibitem[{{Bayer} et~al.(2013){Bayer}, {Bürger}, and {Allgöwer}}]{Bayer2013}
{Bayer}, F., {Bürger}, M., and {Allgöwer}, F. (2013).
\newblock Discrete-time incremental {ISS}: A framework for robust {NMPC}.
\newblock In \emph{Proc European Control Conf. (ECC)}, 2068--2073.

\bibitem[{{Bayer} et~al.(2016){Bayer}, {M{\"u}ller}, and
  {Allg{\"o}wer}}]{Bayer2016b}
{Bayer}, F.A., {M{\"u}ller}, M.A., and {Allg{\"o}wer}, F. (2016).
\newblock Min-max economic model predictive control approaches with guaranteed
  performance.
\newblock In \emph{Proc IEEE 55th Conf. on Decision and Control (CDC)},
  3210--3215.

\bibitem[{Bayer(2017)}]{Bayer2017}
Bayer, F.A. (2017).
\newblock \emph{Performance and Constraint Satisfaction in Robust Economic
  Model Predictive Control}.
\newblock Logos Verlag Berlin GmbH.
\newblock PhD Thesis.

\bibitem[{Bayer et~al.(2014)Bayer, M{\"u}ller, and Allg{\"o}wer}]{Bayer2014}
Bayer, F.A., M{\"u}ller, M.A., and Allg{\"o}wer, F. (2014).
\newblock Tube-based robust economic model predictive control.
\newblock \emph{Journal of Process Control}, 24(8), 1237 -- 1246.

\bibitem[{Bayer et~al.(2018)Bayer, M{\"u}ller, and Allg{\"o}wer}]{Bayer2018}
Bayer, F.A., M{\"u}ller, M.A., and Allg{\"o}wer, F. (2018).
\newblock On optimal system operation in robust economic {MPC}.
\newblock \emph{Automatica}, 88, 98 -- 106.

\bibitem[{{Dong} and {Angeli}(2018)}]{Dong2018}
{Dong}, Z. and {Angeli}, D. (2018).
\newblock Tube-based robust economic model predictive control on dissipative
  systems with generalized optimal regimes of operation.
\newblock In \emph{Proc IEEE Conf. on Decision and Control (CDC)}, 4309--4314.

\bibitem[{Faulwasser et~al.(2018)Faulwasser, Gr{\"u}ne, and
  M{\"u}ller}]{Faulwasser2018}
Faulwasser, T., Gr{\"u}ne, L., and M{\"u}ller, M.A. (2018).
\newblock Economic nonlinear model predictive control.
\newblock \emph{Foundations and Trends in Systems and Control}, 5, 1--98.

\bibitem[{Gr{\"u}ne(2013)}]{Gruene2013}
Gr{\"u}ne, L. (2013).
\newblock Economic receding horizon control without terminal constraints.
\newblock \emph{Automatica}, 49(3), 725 -- 734.

\bibitem[{Gr{\"u}ne and Stieler(2014)}]{Gruene2014}
Gr{\"u}ne, L. and Stieler, M. (2014).
\newblock Asymptotic stability and transient optimality of economic {MPC}
  without terminal conditions.
\newblock \emph{Journal of Process Control}, 24(8), 1187 -- 1196.

\bibitem[{Mayne et~al.(2005)Mayne, Seron, and Raković}]{Mayne2005}
Mayne, D., Seron, M., and Raković, S. (2005).
\newblock Robust model predictive control of constrained linear systems with
  bounded disturbances.
\newblock \emph{Automatica}, 41(2), 219 -- 224.

\bibitem[{{Olshina} et~al.(2018){Olshina}, {Manzie}, {Hield}, and
  {Brear}}]{Olshina2018}
{Olshina}, N., {Manzie}, C., {Hield}, P., and {Brear}, M. (2018).
\newblock A framework for robust nonlinear economic {MPC} without terminal
  constraints for a class of time varying systems.
\newblock In \emph{Proc 15th International Conf. on Control, Automation,
  Robotics and Vision (ICARCV)}, 668--673.

\end{thebibliography}

\end{document}